# TechNet: Technology Semantic Network Based on Patent Data

Serhad Sarica

serhad_sarica@mymail.sutd.edu.sg

Jianxi Luo

luo@sutd.edu.sg

Kristin L. Wood

kristinwood@sutd.edu.sg

**Abstract**

The growing developments in general semantic networks, knowledge graphs and ontology databases have motivated us to build a large-scale comprehensive semantic network of technology-related data for engineering knowledge discovery, technology search and retrieval, and artificial intelligence for engineering design and innovation. Specially, we constructed a technology semantic network (TechNet) that covers the elemental concepts in all domains of technology and their semantic associations by mining the complete U.S. patent database from 1976. To derive the TechNet, natural language processing techniques were utilized to extract terms from massive patent texts and recent word embedding algorithms were employed to vectorize such terms and establish their semantic relationships. We report and evaluate the TechNet for retrieving terms and their pairwise relevance that is meaningful from a technology and engineering design perspective. The TechNet may serve as an infrastructure to support a wide range of applications, e.g., technical text summaries, search query predictions, relational knowledge discovery, and design ideation support, in the context of engineering and technology, and complement or enrich existing semantic databases. To enable such applications, the TechNet is made public via an online interface and APIs for public users to retrieve technology-related terms and their relevancies.

*Keywords:* knowledge discovery; word embedding; technology semantic network; knowledge representation

# 1. Introduction

A large semantic network is normally composed of a library of semantic entities (e.g., words or phrases) and their semantic relations, which are often statistically or linguistically "learned" based on collaboratively edited and accumulated knowledge databases, such as Wikipedia. Over the past decade, semantic networks have been enabled by the development of large-scale knowledge basis and ontology databases, such as WordNet (Fellbaum, 2012; Miller et al., 1990), ConceptNet (Speer et al., 2016; Speer & Havasi, 2012; Speer & Lowry-Duda, 2017), never-ending language learning (NeLL) (Mitchell et al., 2015), Freebase (Bollacker et al., 2008, 2007) and Yago (Rebele et al., 2016), for various general applications in text data mining, natural language processing (NLP), knowledge discovery, information retrieval and artificial intelligence. Likewise, the proprietary Google Knowledge Graph[1] provides the backbone behind Google's semantic search and answer features for web searches, Gmail, and Google Assistant, for example. IBM Watson integrates and utilizes various public semantic networks to identify the most meaningful answers to natural language questions.

Inspired by the growing application of these general semantic networks, we aim to build a technology-focused semantic network to meet the growing demands for engineering knowledge discovery, technology information retrieval, engineering design aids and innovation management. Such a semantic network needs to contain terms that represent a wide variety of technological concepts and their semantic relations that are established by processing the data for engineering designs and technologies. That is, our interest is a "ConceptNet" built on technology-related data for engineering or technology intelligence in general. Hereafter, we call it the "Technology Semantic Network (TechNet)". In turn, the TechNet would support technology-related data integration, knowledge discovery and in-depth analysis at the semantic level (rather the document level) and serve as critical infrastructure for future developments of artificial intelligence in and for engineering and technology innovation.

In a recent effort, Shi et al. (2017) fetched and analysed nearly one million engineering papers from ScienceDirect since 1997 and one thousand design posts from blogs and design websites, such as Yanko Design, to create a large semantic network in the engineering and design contexts. Although our interests are aligned, it is unclear if their mixed data of different types; e.g., academic papers and design blogs, and the sole ScienceDirect publication data (for 20 years) can provide inclusive and balanced representation of engineering knowledge in different domains. In contrast to academic papers and design blogs, patent documents contain technical descriptions of products and processes and are externally validated through relatively objective examinations to ensure usefulness and novelty, and thus naturally avoid data redundancies. Particularly, the USPTO patent database has appeared as the most detailed and comprehensive digital data source about engineering designs in human history and continues to grow organically as inventors file patent applications over time. Thus, it presents a natural choice for the construction of the TechNet.

---

[1] https://googleblog.blogspot.com/2012/05/introducing-knowledge-graph-things-not.html

In the literature, various text analysis methods and tools have been developed to retrieve design information and discover patents for design support or intellectual property management. For instance, Fu et al. (2013) associated patents in a Bayesian network based on the latent semantic analysis of text from different patents to support the retrieval of patents for design inspiration. Mukherjea et al. (2005) created the BioMedical Patent Semantic Web of biological terms within biomedical patent abstracts and their semantic associations to support patent infringement assessments. Chau et al. (2006) employed semantic analysis on nanoscale science and engineering patents to create self-organizing maps of the field. Most patent text analyses have been limited to small samples of patent documents or by the retrieval of documents. A few recent studies utilized the complete patent database to construct large network maps of all patent classes in the classification system according to patent co-classification or co-citations to approximate the total technology space (Alstott et al., 2017; Kay et al., 2014; Yan & Luo, 2017). These networks, despite covering all possible domains of known technologies, are based on the existing patent classification system and document-level analyses.

To construct the TechNet based on engineering data for engineering, our approach is to 1) mine the complete USPTO granted patent database to date, 2) utilize NLP techniques to extract generic engineering terms (which represent functions, components, structures, mechanisms and working principles) from raw patent texts, and 3) use neural network-based word embedding models to vectorize the engineering terms and establish their relationships in a unified vector space to form the technology semantic network and represent the total engineering knowledge base. The mining of the total patent database is aimed to ensure comprehensive coverage of knowledge in all engineering and technology domains. Word embedding algorithms are expected to derive the term vectors and their associations in the vector space to form the semantic network in a data-driven and unsupervised manner, in contrast to the human or social construction of prior semantic networks (e.g., WordNet, ConceptNet). This approach is not trivial due to the massive size and complexity of the patent text database, the uncertainty in the NLP settings, and the lack of benchmark tasks and datasets to assess the resultant alternative networks.

The paper is organized as following. Section 2 provides a literature review on relevant NLP techniques and especially the word embedding models. We will describe our procedure to construct the TechNet in detail in section 3 and then evaluate the resulting alternative TechNet in comparison with existing semantic networks in section 4. We have further developed a web-based interface (www.Tech-Net.org) for public users to retrieve technology-related terms and their semantic associations, which will be presented in Section 5.

## 2. Related Work

Our research is inspired by the prior works on semantic representations and organizations of information and knowledge and aims to utilize the recent word embeddings models for efficiently training the distributed representation of engineering knowledge from large and raw technical language data sources. Therefore, we review the relevant NLP literature in the following subsections.

### 2.1 Semantic representation of information

The emergence of NLP techniques enabled the retrieval of semantic-level information from massive unstructured textual data and the extraction of relational information between semantic units. The relational information can be utilized as an inspiration source in engineering design ideation practices and can be used to expand design-related information queries. More importantly, these techniques can be utilized to capture design-related information and represent them in a structured medium to aid designers to explore the technology knowledge space. These structured media are usually called ontologies, which serve as a hierarchy, use a lexicon and construct relations within this lexicon. The ontology-based databases retrieve the unstructured data and relate them using various techniques.

However, none of the publicly available ontology-based databases, such as the popular WordNet (Fellbaum, 2012; Miller et al., 1990), ConceptNet (Speer et al., 2016; Speer & Havasi, 2012; Speer & Lowry-Duda, 2017) or BabelNet (Navigli & Ponzetto, 2012), derive the relations between entities from an engineering design viewpoint.

The relations among the entities in ontology-based databases may be hand-built, built using semi-supervised procedures or constructed automatically by utilizing unsupervised methods. As the most popular and indisputably the largest among the hand-built ontologies, WordNet completely relies on experts to retrieve knowledge and relations such as synonymy, hypernymy, hyponymy, and antonymy and required a large amount of human effort and time to reach the current state. The effort can be traced back to mid-1980s. Since generally both the time and the human resources are scarce, hand-built ontologies are mostly domain-specific (Ahmed et al., 2007; Z. Li et al., 2008, 2009) instead of WordNet. On the other hand, the availability of basic NLP tools led to the introduction of various semi-automatic ontology generation methods. For instance, Reinberger et al. (2004) extracted terms from a corpus group based on their semantic relations, which could then be refined by a domain expert. Nobécourt (2000) introduced a method that expects domain experts to evaluate the extracted keywords to generate the skeleton of the ontology. Alfonseca & Manadhar (2002) modelled a concept using its context by means of co-occurrence statistics and used distance metrics to quantify semantic relations.

Along with the methodology development is the emergence of several large public ontology databases constructed on public online data in the past decade. BabelNet (Navigli & Ponzetto, 2012) introduces an automatic framework, by integrating Wikipedia entries and WordNet, to create a large multi-lingual semantic knowledge database that constructs a graph structure with various kinds of links between entities such as is-a, part-of, etc. ConceptNet (Speer et al., 2016; Speer & Havasi, 2012; Speer & Lowry-Duda, 2017) utilizes unsupervised and semi-supervised methods to retrieve the knowledge from internet resources, such as IsA, MadeOf, and PartOf. Freebase (Bollacker et al., 2008, 2007), later acquired by Google and dissolved in Google Knowledge Graph, mines resources such as Wikipedia and enables a collaborative environment to handle the organization, representation, and integration of large and diverse data sets, thus facilitating continuous growth. On the other hand, Yago (Rebele et al., 2016) follows the Resource Description Framework (RDF) (World Wide Web Consortium, 2014) triplets to store relational data mined from Wikipedia, WordNet, and GeoNames and filtered to fit predefined relational structures.

The studies to retrieve entities and relational knowledge in engineering and technology intelligence fall into two categories: document retrieval and concept retrieval. Chakrabarti et al. (2006) introduced a design repository where design documents are represented with a Function-Behaviour-Structure (FBS) model in a machine comprehensible format, letting users run queries based on the semantic similarity of the words forming the FBS model. Kim & Kim (2012) mines the causal functions and effect functions as well as the objects related to these functions from patent texts to construct a function network to enable the search for analogical inventions. Murphy et al. (2014) introduced a method for querying functional analogies by representing documents using functional verbs and mapping them to a vector space model. Sosa et al. (2014) introduced a semantic-based approach to explore the design documents for reconfigurable or transformable robotics. They used WordNet to form a lexically hierarchical structure with abstracted functional verbs. Glier et al. (2018) used a bag-of-words (BOW) method with stemmed words to represent the documents with vectors for the search in a database of bioinspired design documents. Djenouri et al. (2018) represented documents with BOW in the vector space, clustered them with the k-means algorithm, and used bee swarm optimization to retrieve the documents.

Aside from the studies of document retrievals, Li et al. (2005) utilized basic NLP techniques and semantic analysis to mine relational engineering design knowledge and map them to a predefined ontological tree to generate a domain-specific engineering design ontology. Ahmed et al. (2007) proposed a methodology to create engineering design ontologies where the reuse of taxonomies for general engineering concepts is favoured, while product-related taxonomies and relational knowledge are built based on expert knowledge. Li et al. (2008, 2009) introduced a process of creating engineering ontologies for a predefined engineering domain that populates the ontology and derives the relations among entities using worksheets in a machine-readable format produced by human contributors. Jursic et al. (2012) used a predefined term set in a bisociative information network and formed a term-document matrix by mining the text in a specific medical field to acquire the relational knowledge between concepts. Tan et al. (2016) introduced a semi-supervised method to retrieve the semantic units and relations among them in a specific domain by training classifiers with a limited number of structured-unstructured textual data pairs. Gutiérrez et al. (2016) introduced a framework to integrate the information in various structured knowledge sources aligned with WordNet, such as WordNet Affect (Strapparava & Valitutti, 2004), WordNet Domains (Magnini & Cavagli, 2000) and Suggested Upper Merged Ontology (Niles & Pease, 2003) to better measure the similarity between user comments in online platforms. Munoz & Tucker (2018), with a focus on the context for the semantic meaning of terms and instead of a classical BOW approach, created a semantic network of terms by quantifying the relations based on co-occurrence within a predefined context window. Shi et al. (2017) introduced an unsupervised process to retrieve concepts and learn the relations among concepts gradually while parsing the design-related textual information. The concepts are related to each other if they appear in the same sentence. Martinez-Rodriguez et al. (2018) proposed a knowledge graph construction method that utilizes text processing methods to find named entities and obtain binary relations from open knowledge sources. Barba-Gonzalez et al. (2019) introduced a hierarchically structured domain-specific ontology by inheriting the knowledge of readily available domain-specific ontologies and extending these works with the help of domain experts.

The capabilities and focuses of the prior approaches to semantic networks are summarized in Table 1, with highlighting if they are engineering-related, hand-built and unsupervised.

**2.2 Word embeddings models**

Word embeddings is the general name of a set of NLP techniques and methods that represent words and phrases in a raw text as dense vectors of real numbers. The word vectors have a wide range of applications in information retrieval, machine translation, document classification, question answering, named entity recognition, and parsing (Pennington et al., 2014). In classical approaches such as BOW, a sparse vector represents each of the words/phrases in a text whose dimension is the total number of unique words/phrases. Word embedding-based methods transform this high dimensional space of words to a relatively low dimensional space. word2vec (Mikolov, Corrado, et al., 2013) and GloVe (Pennington et al., 2014) are two recent and popular word embedding algorithms used in real-world applications as well as in academic literature. Both algorithms consider the relationships of words with their contexts while training word embeddings. For example, in a general knowledge corpus such as Wikipedia, the (computer, software) pair would appear more frequently than (computer, food) pair. In addition, the context of "computer" would be more similar to the context of "software" compared to "food". Consequently, the word embedding of "computer" would be more similar to the word embedding of "software" than "food".

Table 1: Previous studies on information and semantic relation retrieval. (+: "partially satisfied")

| | Engineering Related | Hand built | Un-supervised | Unit of Interest | NLP Processes | Statistical Basis | Training | Relations |
|---|---|---|---|---|---|---|---|---|
| **WordNet** | | x | | words or phrases | | | | synonymy, hyponymy, meronomy, troponymy, antonymy |
| **ConceptNet** | | + | + | words or phrases | | | | 36 different binary relations |
| **ConceptNet Numberbatch** | | | x | words or phrases | Filtering, merging readily available sources | | Retrofitting (Dyer et al., 2014) | cosine similarity |
| **Prebuilt word2vec** | | | x | words or phrases | Tokenize, phrase chunking | co-occurrence in a context window | Modified NNLM | cosine similarity |
| **Prebuilt GloVe** | | | x | words or phrases | Tokenize, phrase chunking, co-occurrence statistics | global co-occurrence frequencies in a context window | Log-bilinear regression | cosine similarity |
| **Juršic et al. (2012)** | x | | x | words or phrases | Tokenize, filter, stem, entity extraction, tf-idf | Vector space model (BOW) | | cosine similarity |
| **Murphy et al. (2014)** | x | | x | document | Tokenize, stem, filter | Vector space model (BOW) | | cosine similarity |
| **Z. Li et al. (2005)** | x | | + | words or phrases | Tokenize, filter, POS Tagging, bracketing (Glasgow et al., 1998) | | | 8 different binary relations |
| **Z. Li et al. (2009)** | x | + | + | words or phrases | | | | 12 different relations quantified between 0-1 |
| **Glier et al. (2018)** | x | | x | document | Tokenize, Stem, Filter, Information Gain | | Naïve Bayes | Document Classification |
| **Munoz & Tucker (2018)** | x | | x | document | Tokenize, Filter | co-occurrence in a context window | | Network clustering |
| **Shi et al. (2017)** | x | | x | words or phrases | POS Tagging, phrase chunking | Probability and velocity analysis | | normalized network distance |
| **Our approach** | x | | x | words or phrases | Tokenize, phrase detection, filter, lemmatization | co-occurrence in a context window | GloVe / word2vec | cosine distance |

Unlike the well-known Latent Semantic Analysis (LSA) and Latent Dirichlet Allocation (LDA), which also focus on estimating continuous representations of words, word2vec uses an artificial neural network to derive the continuous representations of words and was shown to perform better than LSA for preserving linear regularities among words (e.g., the relations between family-families≈car-cars (Mikolov, Yih, et al., 2013)) and have greater computational efficiency than LDA when the data set becomes large (Mikolov, Corrado, et al., 2013). Mikolov, Corrado, et al. (2013) named the network architecture "Continuous Skip-Gram" since the training process skips the target word and uses a predefined window of neighbouring words before and after the target word as the context. The model classifies the surrounding words in the context window of the target word using a softmax classifier by minimizing the classification error. The overview of the model's network architecture is illustrated in Fig. 1 where a simple training instance is illustrated.

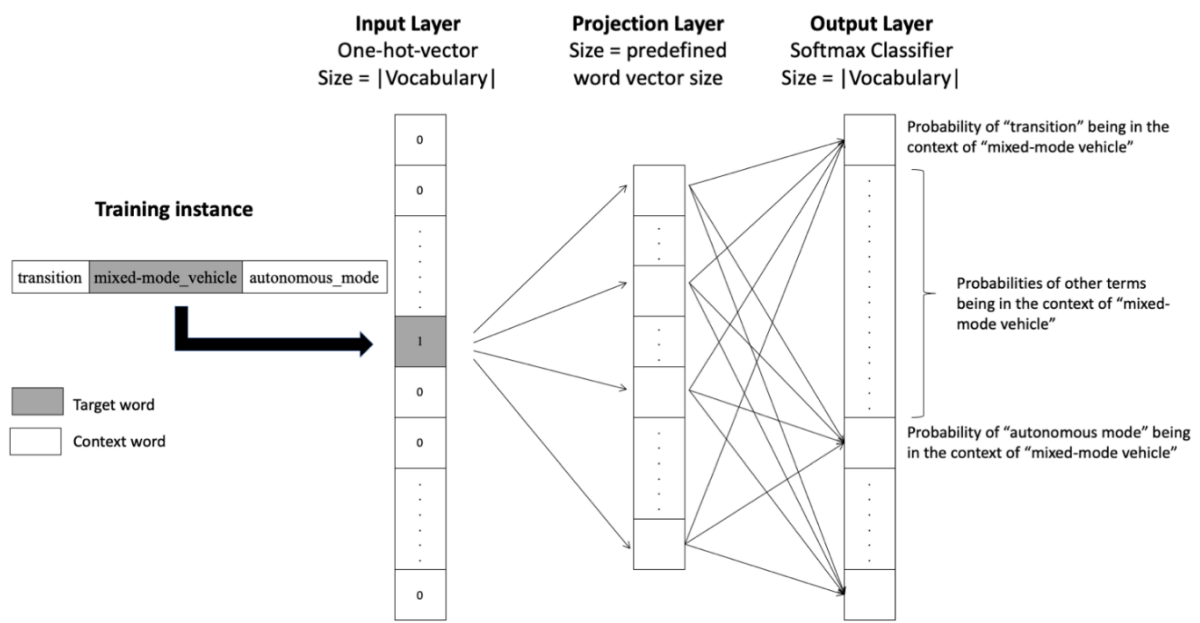

Fig. 1. Overview of the skip-gram network architecture with a window size of 1.

As seen in Fig. 1, the neural network model has 3 layers, where each word is represented by a one-hot vector in the input layer. The size of these input vectors is equal to the size of the vocabulary; i.e., the number of unique terms in the training material. The projection layer is a non-activated hidden layer that directly acts as a word vector lookup table. Thus, together with the input layer, the projection layer directly maps the word embedding of the target word. The output layer's size is also equal to the size of the vocabulary and implements a softmax regression classifier to calculate the probabilities of each term being in the context of the target word. This architecture differs from the typical neural net language model (NNLM) architecture (Bengio et al., 2003) in that it does not have a non-linear hidden layer and thus diminishes the effect of words in history on the projection layer in the NNLM.

On the other hand, GloVe (Pennington et al., 2014) seeks to capture the embedding of a single word within the overall structure of the corpus (i.e., the entire corpus of sentences) and concerns the global co-occurrence counts of words. Similar to word2vec, GloVe also assumes that the probability of co-occurrences of contextually close words is much higher than that of contextually irrelevant words, and uses a context window to train word embeddings. word2vec's and GloVe's ability to easily handle large amounts of data in an efficient way and their ability to capture non-obvious and indirect relations between terms make them promising candidates for the retrieval of technical terms and relations among them in the large patent text database.

In sum, most of the large ontological databases lack a sufficiently representative technology-related terminology and engineering design-related viewpoint on relations among the semantic units. On the other hand, engineering design-related studies generally focus on small sets of data or a specific technical field, which limits the possible expansion of the studies to cover a wide variety of engineering domains. To date, the patent database that covers all domains of technology (World Trade Organization, 1995) has not been utilized to build a large-scale and comprehensive knowledge graph by retrieving engineering and technology-related concepts and relations among them from patent texts. This study fills the gap. Patent texts contain rich design information and the patent database covers all domains of technologies and engineering practices. Specifically, we extract technology-related terms in patent documents by using NLP techniques, and mine relations among them by utilizing word-embedding algorithms that focus on relations among terms and their contexts. The resultant TechNet is unique since it directly represents the relations of semantic units, which represent technical functions, components, structures, properties, operating principles or other concepts.

## 3. Construction of the TechNet

Fig. 2 depicts the overall framework of our methodology to build the TechNet. The key steps include the extraction of technically meaningful terms (words or phrases) from patent texts and the use of word embedding models to derive the vector representations of these terms, which form a vector space and can be further associated to forge the semantic knowledge network of technological concepts.

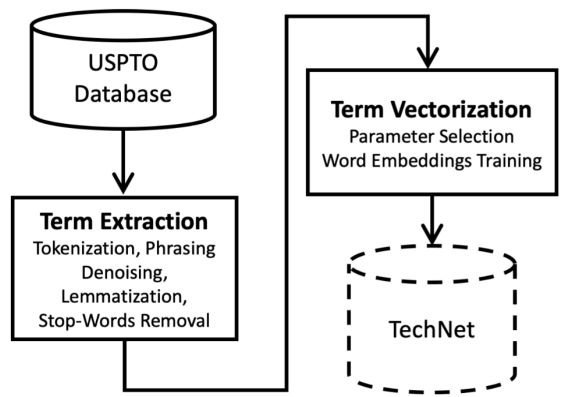

Fig. 2: Overall Framework

**3.1 Source Data**

The design repositories contain and organize information of prior art, design, and technologies, which can enable designers to combine, recombine, transfer, adapt or adopt this information in actual design practices (Bohm et al., 2008). Unlike the manually built design repositories, the patent database contains a naturally built systematic catalogue of the technologies invented so far, and it expands with time as new technologies are introduced by inventors to protect their rights. Moreover, patents contain significant information on a wide variety of technologies and in a broad range of engineering domains. The unstructured information in the patents also presents the building blocks of technologies and their relations with each other. In this study, the patent database is utilized to retrieve the technology-related terms and associations between them.

Our analysis utilizes the complete USPTO patent database containing 6.3 million documents for the patents granted from 1976 to October 2017 (access date: 20/10/2017)[2]. We chose to analyse only utility patents that are about technologies and excluded the design patents for look-and-feel aesthetics designs, for our focus on technology-related knowledge retrieval. Therefore, the design patents are filtered out, returning 5,771,030 patents for our use. Moreover, we chose to focus on titles and abstracts in patent documents to ensure computational efficiency and accuracy for the retrieval of technology-related knowledge from millions of patent documents. The title and abstract of a patent document provide the most accurate and concise description of a patented invention because they must contain all the necessary terms to explain the invention and must not contain texts on other information than the technology itself. By contrast, legal claims are developed by lawyers and use disguised terms to cover more areas than the inventive technology itself for legal defensiveness and strategic reasons. The technical description also contains broader content about contexts, backgrounds and other technologies than the technology itself. Such broad texts introduce noise and reduce the accuracy of the statistical procedure to extract the abstract knowledge presentation of the inventive technology.

---

[2] An up-to-date version of the database can be downloaded from http://www.patentsview.org/download/

**3.2 Term Extraction**

The collection of patent titles and abstracts is first transformed into a line-sentence format, utilizing the sentence tokenization method in the Natural Language Toolkit (NLTK)[3]. All the text in the corpus is normalized to lowercase letters to avoid additional vocabulary caused by lowercase/uppercase differences in the same words. The punctuation marks in sentences are removed except "-" and "/". These two special characters are frequently used in word-tuples, such as "AC/DC" and "inter-link", which can be regarded as a single term. Stop-word removal is postponed after phrase detection because some of the phrases contain stop-words. These basic pre-processing steps transferred the original raw texts into a collection of 26,756,162 sentences, including approximately 699 million words.

Then, we used various algorithms and methods to identify phrases in the collection of sentences and evaluated their performances on the retrieval of technology-related phrases. First of them was from Mikolov et al. (2013). The algorithm finds words that frequently appear together, and in other contexts infrequently, by using a simple statistical method based on the count of words to give a score to each bigram such that:

$$score(w_i, w_j) = \frac{(count(w_i w_j) - \delta) * |vocabulary|}{count(w_i) * count(w_j)} \quad (1)$$

where $count(w_i w_j)$ is the count of $w_i$ and $w_j$ appearing together as bigrams in the collection of sentences and $count(w_i)$ is the count of $w_i$ in the collection of sentences. The parameter $\delta$ is used as a discounting coefficient to prevent too many phrases consisting of very infrequent words from being formed. We set $\delta = 2$ to prevent the algorithm from detecting phrases, which includes pairs of words that co-occur less than twice. The term |vocabulary| denotes the size of the vocabulary. Bigrams with a score over a defined threshold ($T_{phrase}$) are used as phrases and joined with a "_" character in the corpus, to be treated as a single term. We run the phrasing algorithm of Mikolov et al. (2013) on the pre-processed corpus twice to detect *n*-grams, where *n* = [2,4]. The first run detects only bigrams by employing a higher threshold value $T_{phrase}^1$, while the second run can detect *n*-grams up to *n* = 4 by using a lower threshold value $T_{phrase}^2$ to enable combinations of bigrams. Via this procedure of repeating the phrasing process with decreasing threshold values of $T_{phrase}$, we detected phrases that appear more frequently in the first step using the higher threshold value, e.g., "autonomous vehicle", and detected phrases that are comparatively less frequent in the second step using the lower threshold value, e.g., "autonomous vehicle platooning". Three different sets of threshold tuples ($T_{phrase}^1$, $T_{phrase}^2$), specifically (200, 100), (50, 25), (5, 2.5) (one is comparatively high, one is very low, and the other one is in between), were used to detect phrases with different sensitivity levels.

---

[3] https://www.nltk.org/. NLTK is a suite of libraries and programs for NLP using the Python programming language.

As the second algorithm, we used a simple rule-based noun phrase (i.e., noun+noun and adj+noun) extractor by training a tagger in the NLTK tool on part-of-speech (POS) tagged Brown Corpus (Francis & Kucera, 1964). As the third, we used the TextRank algorithm (Mihalcea & Tarau, 2004), a graph-based unsupervised extraction method. TextRank constructs a network of words based on collocations of the words in the text, and scores them by measuring their importance in this network structure, followed by merging top-ranking words as phrases if they are collocated in the text. For TextRank, we used a co-occurrence relation of 3, which defines the maximum distance between two words in a sentence to be connected in the network structure. As the last, we employed the Rake algorithm (Rose et al., 2010), another graph-based unsupervised extraction method. Rake generates a candidate keyword list by merging the words if they are delimited by a stop-word and construct a co-occurrence network by using these candidate keywords to score the words. Based on the word scores, candidate keyword scores are calculated, and the top-ranking phrases are elected to be merged. All these algorithms are tailored so that they can return phrases up to four words long. Table 2 reports the vocabulary size and number of phrases resulting from each phrasing method. All of the algorithms we have used are allowed to form phrases up to 4-grams.

Table 2: Statistics for the corpora after phrasing process

| Corpus ID | Phrasing Algorithm | Vocabulary Size | Number of Phrases | Processing Time (mins)* |
|---|---|---|---|---|
| 1 | Mikolov, et al., (2013) ($T_{phrase}^1$=200 $T_{phrase}^2$=100) | 3,241,111 | 1,535,617 | 25.15 |
| 2 | Mikolov, et al., (2013) ($T_{phrase}^1$=50 $T_{phrase}^2$=25) | 3,824,388 | 2,122,495 | 27.01 |
| 3 | Mikolov, et al., (2013) ($T_{phrase}^1$=5 $T_{phrase}^2$=2.5) | 14,173,083 | 12,482,611 | 21.75 |
| 4 | Rule-based | 7,663,849 | 5,585,080 | 470.85 |
| 5 | TextRank (Mihalcea & Tarau, 2004) | 2,993,711 | 545,392 | 827.21 |
| 6 | Rake (Rose et al., 2010) | 11,605,866 | 8,278,176 | 220.46 |

*The processing was held in a computer with 3.8GHz processor and 16GB of RAM using single core

After the phrase detection process, we continued to further clean the data. During the phrasing process, especially for the work of Mikolov et al. (2013), some noisy phrases are formed due to their statistical significance, such as phrases including stop words such as "the_", "a_", and "and_". A custom filter of the noisy terms and phrases was built with the help of a human reader who is a researcher in patent document analysis and familiar with patent jargon. The human expert read 1,000 randomly selected sentences from the corpus having the most populated collection of phrases to detect noisy phrase formation patterns and stop-words that need to be removed. Although readily built stop-word lists were also utilized in this study, the reader additionally detected patent specific stop-words (e.g., disclosure, plurality, thereof) in addition to the obvious ones.

Next, all the words are represented with their regularized forms to avoid having multiple terms representing the same word or phrase and thus decrease the vocabulary size. This step is achieved by first using a POS tagger (Toutanova & Manning, 2007) to detect the type of words in the sentences and lemmatize those words accordingly. For example, if the word "learning" is tagged as a VERB, it would be regularized as "learn" while it would be regularized as "learning" if it is tagged as a NOUN. Then, we remove the stop-words, which are meaningless, but their frequent occurrences in the database can distort frequency-based statistical analyses. Specifically, we removed the stop-words in NLTK's English stop-word list, USPTO's patent stop words list ("Stopwords, USPTO Full-Text Database," n.d.) and the previously detected unconventional stop-words list. Words or the parts of phrases containing only digits are also removed. Finally, we also filtered out the terms appearing only once because these rarely occurring terms are likely to be misspelt words, nonsense words, or insignificant ones. The overall phrase detection and denoising procedure is detailed in Fig. 3.

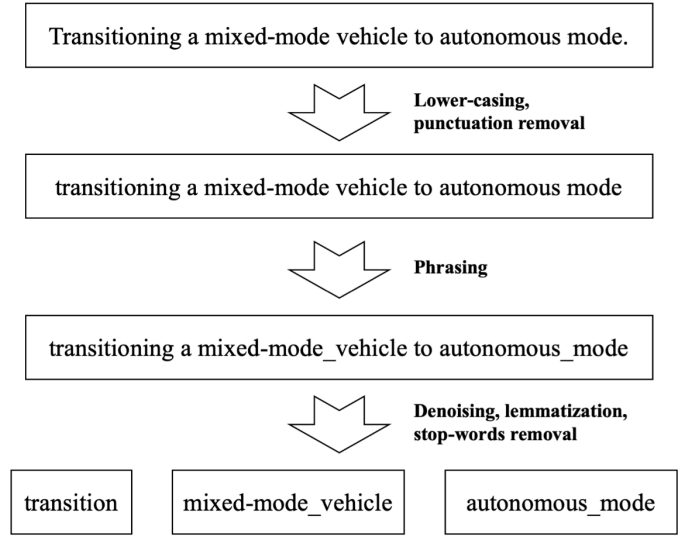

Fig. 3: The steps to derive term tokens, using the title of patent US8078349 as an example

Finally, in our database, each sentence is represented as a sequence of term tokens that can be either words or phrases, as illustrated in Fig. 4. From now on, we refer to the total collection of tokenized sentences as a "corpus". Table 3 presents the count statistics of the same set of six corpora reported in Table 2. The vocabularies of these corpora differentiate themselves from WordNet, ConceptNet and others in that they are based on patent data and specialized in technology-related terms.

Table 3: Statistics for the corpora after denoising and lemmatization

| Corpus ID | Number of Unique Terms (i.e., the Vocabulary Size) | Number of Phrases |
|---|---|---|
| 1 | 1,593,465 | 795,482 |
| 2 | 1,888,963 | 1,090,015 |
| 3 | 4,038,924 | 3,233,852 |
| 4 | 5,714,472 | 5,037,313 |
| 5 | 1,514,565 | 717,811 |
| 6 | 6,528,322 | 5,942,049 |

**3.3 Term Vectorization**

A pre-processed corpus is then used to derive vector representations of the terms in an unsupervised manner using word embedding algorithms. We experimented with both Word2vec and GloVe algorithms for the word embedding training. The training process additionally requires setting the values for the hyperparameters in such algorithms, particularly the vector size and context window size. In the literature on word-embeddings models (Elekes et al., 2017; Mikolov, Chen, et al., 2013; Mikolov, Corrado, et al., 2013; Pennington et al., 2014) and applications (Banerjee et al., 2018; Kuzi et al., 2016; Li et al., 2018; Risch & Krestel, 2019), researchers normally experimented different values for such parameters and determined the values according to specific contexts and needs.

The first parameter (i.e., vector size) defines the size of the word vectors to be calculated by the word-embedding model and is the size of the projection layer of the neural networks to be trained. Practical studies on word-embedding models generally vary vector sizes, starting from a comparatively short vector as short as 50, and ending with a vector size as long as 1000. In this study, we chose not to start with a very small-sized vector taking into account the size of the vocabulary and not to end with a very large vector which could reduce practical efficiency of possible applications due to computational complexity. Regarding the vector size, we experimented with the values of 150, 300 and 600.

The second parameter is *context window size* and defines the context window size right and left to the target term. We take into account the sentence lengths in the corpus to determine the window sizes for training. Fig. 4 plots the cumulative distribution of sentences by length for the corpus #3 and suggests that using a window size of 10 (10 context words to left + target word + 10 context words to right) guarantees that the whole sentence of length 20 will be treated as context at least once while training terms for more than 90% of sentences. On the other hand, using a window size of 20 guarantees that the whole sentence will be treated as context all the time while training the words in them for more than 90% of sentences. Therefore, we experimented with both window size parameters (10 and 20).

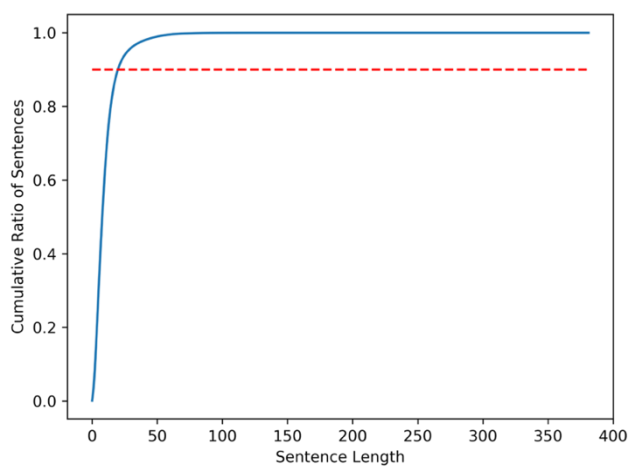

Fig. 4. Normalized cumulative distribution of sentences by length for corpus #3. The dotted red line represents a threshold of 0.9 and intersects the curve at a sentence length of 20.

In addition, the word embedding models also allow filtering the terms with an occurrence frequency higher than a "down-sampling" parameter from the training process. The most frequently occurring terms are likely to be mundane and contextually meaningless and can distort the statistics. A term *i* would be ignored with a likelihood of $p_i$ in training if its frequency $f_i$; i.e., the ratio of the count of occurrences of the term to the total number of terms in the corresponding corpus, is higher than the down-sample threshold value *d*, according to the following equation:

$$p_i = 1 - \sqrt{\frac{d}{f_i}} \qquad (2)$$

To determine *d* in equation (2), we check the term frequencies in the corpus. For instance, Fig. 5 reports the frequencies of the 100 most frequent terms in corpus #3 above. The high occurrence frequencies of the top 10 terms make them good candidates to sample down. We further checked the specific top 10 frequent terms (i.e., method, form, least, comprise, system, connect, receive, base, position, control) and found that terms following the term 'system' appear to be meaningful for some technologies. Therefore, we selected the frequency of the following term 'connect' (0.0039) as the down-sampling threshold *d* in equation (2) for calculating the down-sampling probability $p_i$ for each term. The words that are ignored when they appear in the context window during training are still counted as target words and remain in the corpora.

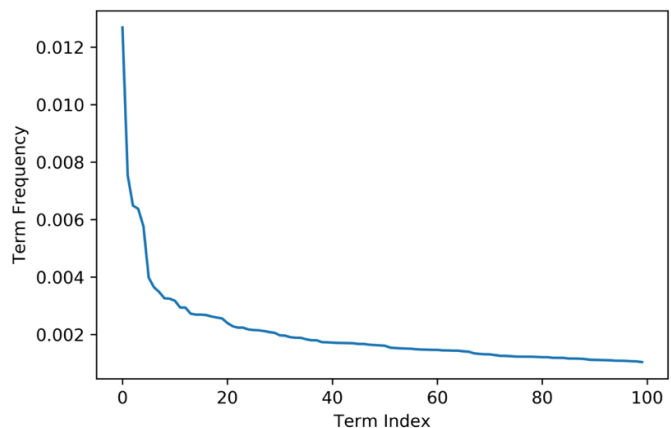

Fig. 5. Top 100 most frequent terms and their term frequencies for corpus #3

By applying word2vec and GloVe algorithms with varying window sizes (small, large) and vector sizes (small, medium, large) to the corpus with the highest term retrieval performance, we trained

12 different sets of term vectors for comparisons later in section 4. The duration of training on the full patent database varied between 527 and 2,164 minutes using a computer with a 3.5 GHz Intel Xeon processor.

After the terms have been represented as vectors, we can associate them by calculating the cosine similarity between these vectors and form a semantic network of technological concepts (TechNet). We consider the angular similarity of the word-embedding vectors of different terms an indicator of their "semantic relevance". For a fully connected undirected network of $n$ terms, there exists $n(n-1)/2$ links. For this study, despite the models and parameters, the corpora contain a few millions of unique terms and therefore more than $10^{12}$ potential bidirectional links between these terms in the TechNet. Given the large size and connectivity of the TechNet, we only store the vector representations of these terms and conduct on-demand retrieval of the semantic relevance between terms.

## 4. Evaluation

We evaluated the efficacy of the TechNets (arising from different phrase detection techniques, different word embedding models and hyperparameters) for knowledge retrieval and inference. Thus, the evaluation is two-fold. The first is to evaluate the engineering-related term coverage of a semantic network for engineering concept retrieval. The second is with regard to making knowledge inference and reasoning from one term to another according to their semantic relevance in the semantic network. In particular, the evaluation needs to be contextualized in engineering and technology.

### 4.1 Evaluation of Term Coverage

To evaluate technical term retrieval performance, we adopted the Multidisciplinary Design Project Engineering Dictionary (Cambridge-MIT Institute Multidisciplinary Design Project, 2006) developed by the Cambridge-MIT Institute at University of Cambridge as the benchmark. This dictionary serves as a general engineering glossary and contains 2,704 terms (including abbreviations) in 6 main categories, namely civil & structural engineering (89 terms), materials engineering (264 terms), mechanical engineering (209 terms), mining engineering (368 terms), nuclear engineering (374 terms) and computer & software engineering (1,400 terms). We compared our six vocabularies, Corpus #1 to #6 (please refer to Table 2 and Table 3 for the details), with those of WordNet[4], ConceptNet[5], the vocabulary of the pre-trained word2vec word vectors[6] based on Google News (3 million words) (Mikolov et al., 2013), the vocabulary of the pre-trained GloVe word vectors[7] based on Wikipedia and Gigaword (400 thousand words) (Pennington et al., 2014), and the vocabulary in the semantic network of Shi et al. (2017) based on engineering paper publications.

The performance is assessed as the portion of the total 2,704 keywords in the Multidisciplinary Design Project Engineering Dictionary (Cambridge-MIT Institute Multidisciplinary Design Project, 2006) that can be retrieved from different vocabularies, given by:

$$R_r = \frac{n_r}{N} \quad (3)$$

where $R_r$ is the retrieval performance, $n_r$ is the number of retrieved keywords and $N$ is the total number of keywords. As reported in Table 4, the vocabulary obtained from the corpus phrased by the algorithm of Mikolov et al. (2013) with the parameter set ($T_{phrase}^1$=5, $T_{phrase}^2$=2.5), i.e. Corpus #3, performed generally better than all others. In the later analysis, we focus on the models trained on this specific corpus. Generally speaking, the superior engineering term coverage of all our vocabularies suggests the richness of the engineering design information and technological knowledge stored in patent documents, and an advantage of our TechNet for engineering knowledge retrieval.

Table 4. Term retrieval performance. Bold scores represent the best performances.

|  | Civil & Structural | Materials | Mechanical | Mining | Nuclear | Computer & Software | Total |
|---|---|---|---|---|---|---|---|
| WordNet | 0.494 | 0.534 | 0.440 | 0.557 | 0.374 | 0.325 | 0.398 |
| ConceptNet | 0.685 | 0.686 | 0.632 | 0.668 | 0.591 | 0.629 | 0.637 |
| Shi et al. | 0.764 | 0.723 | 0.627 | 0.546 | 0.527 | 0.504 | 0.553 |
| Pretrained w2v | 0.449 | 0.496 | 0.446 | 0.590 | 0.372 | 0.463 | 0.470 |
| Pretrained GloVe | 0.449 | 0.458 | 0.440 | 0.563 | 0.342 | 0.576 | 0.567 |
| Corpus #1 | 0.708 | 0.773 | 0.699 | 0.701 | 0.588 | 0.650 | 0.666 |
| Corpus #2 | 0.719 | 0.784 | 0.713 | 0.709 | 0.599 | 0.653 | 0.672 |
| Corpus #3 | 0.876 | **0.841** | 0.761 | **0.799** | **0.698** | 0.671 | **0.723** |
| Corpus #4 | 0.843 | 0765 | 0.694 | 0.715 | 0.586 | 0.661 | 0.677 |
| Corpus #5 | 0.843 | 0.818 | 0.751 | 0.777 | 0.684 | 0.668 | 0.712 |
| Corpus #6 | **0.910** | 0.837 | **0.766** | 0.788 | 0.695 | **0.672** | 0.722 |

---

[4] Accessed via NLTK
[5] https://github.com/commonsense/conceptnet-numberbatch
[6] https://drive.google.com/file/d/0B7XkCwpI5KDYNlNUTTlSS21pQmM/edit?usp=sharing
[7] http://nlp.stanford.edu/data/glove.6B.zip

**4.2 Evaluation of Term-To-Term Semantic Relevance**

Then, following the literature on word embeddings (Mikolov et al., 2013; Pennington et al., 2014; Speer & Lowry-Duda, 2017) we evaluated the performance of the 12 candidate TechNets (trained using word2vec and GloVe, 2 window sizes, 3 vector sizes, and corpus #3) in retrieving pairwise term relevance against human comprehension, based on three readily available benchmark datasets, and one custom technology term relevance (TTR) dataset. The three readily available datasets are Word Similarity-353 (WS353) (Finkelstein et al., 2002) (353 pairs), Rare Words (RW) (Luong et al., 2013) (2,034 pairs) and Stanford's Contextual Word Similarities (SCWS) (Huang et al., 2012) (2,003 pairs). These datasets consist of word tuples and their corresponding average similarity scores evaluated by human participants in non-technical contexts.

We created the TTR dataset by choosing four easily comprehensible technology and engineering terms from each of the main categories contained in Multidisciplinary Design Project Engineering Dictionary (Cambridge-MIT Institute Multidisciplinary Design Project, 2006). We prepared 276 term pairs representing various degrees of relevance and employed 10 human subjects to estimate the technical relevance of each pair of the terms on a scale from 0 (totally unrelated) to 10 (highly related, synonyms or identical terms) following the techniques in the literature (Huang et al., 2012; Luong et al., 2013). The human subjects are experienced engineers and engineering researchers. We used Cronbach's alpha (Cronbach, 1951) to measure the inter-rater reliability that is 0.88. This high value indicates the independence of the evaluations from the human judges. Thus, we used the average of the 10 human ratings for each term pair as the pair relevance score. As shown in Fig. 6, the average relevance scores for the 276 term pairs in our TTR dataset resembles a normal distribution. The TTR benchmark dataset can be accessed via https://github.com/SerhadS/TechNet, the GitHub Repository for this project.

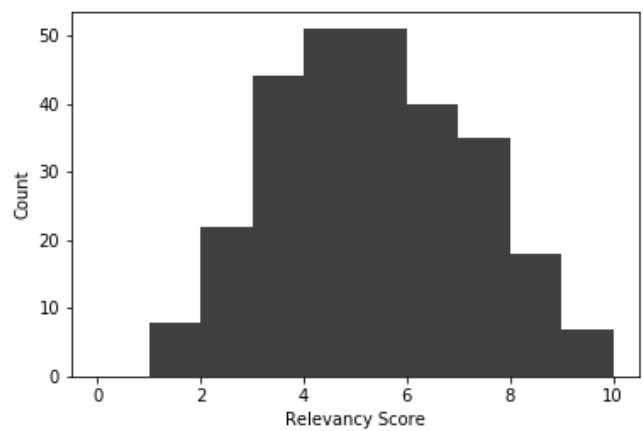

Fig. 6: Histogram of average of evaluator ratings for pairs of terms in TTR

Table 5 reports the Spearman rank correlation coefficients between the pairwise association values of the same term pairs from the four benchmarks and those from our candidate TechNets and other publicly available databases. For the WS353, RW and SCWS benchmark datasets built on general knowledge, ConceptNet, pre-trained Word2Vec and GloVe vectors perform better than the TechNets. This is not surprising because these word embeddings models, knowledge databases and the benchmark datasets were created in the same general-knowledge and non-technical contexts, whereas our TechNets are trained on technical language data and specialized in engineering knowledge inference. It is worth to note that WordNet provides extremely low and much lower correspondence to all four benchmarks than TechNets, despite its popularity in uses to date.

For TechNets alone, they generally present stronger correspondence with the TTR benchmark than the three general-knowledge benchmarks. There is also a clear better performance of the word2vec-trained TechNets than those from GloVe across all four benchmarks. For the TTR benchmark alone, TechNet #9 presents the best correspondence to human evaluations. In fact, the TechNets trained on word2vec generally outperform all other existing semantic networks, although the ones trained on GloVe do not. In brief, our procedure, especially by training the word2vec model on patent data, has generated some semantic networks that outperform existing general-knowledge semantic networks for engineering knowledge inference.

Table 5. Correspondence of various semantic networks with the benchmark datasets. Numbered models are trained on Corpus #3. Bold scores show the best correlations among all models. Underlined scores show the best correlations among our models.

| | | Parameters | | Benchmark Datasets | | | |
|---|---|---|---|---|---|---|---|
| Model | Algorithm | Window size | Vector size | WS353 | RW | SCWS | TTR |
| 1 | word2vec | 10 | 150 | 0.47 | <u>0.31</u> | <u>0.54</u> | 0.63 |
| 2 | GloVe | 10 | 150 | 0.35 | 0.17 | 0.46 | 0.53 |
| 3 | word2vec | 20 | 150 | 0.46 | <u>0.31</u> | 0.52 | 0.62 |
| 4 | GloVe | 20 | 150 | 0.36 | 0.17 | 0.47 | 0.51 |
| 5 | word2vec | 10 | 300 | <u>0.48</u> | 0.30 | 0.53 | 0.64 |
| 6 | GloVe | 10 | 300 | 0.35 | 0.17 | 0.47 | 0.51 |
| 7 | word2vec | 20 | 300 | <u>0.48</u> | 0.28 | 0.52 | 0.63 |
| 8 | GloVe | 20 | 300 | 0.38 | 0.19 | 0.48 | 0.52 |
| 9 | word2vec | 10 | 600 | <u>0.48</u> | 0.28 | 0.52 | **0.66** |
| 10 | GloVe | 10 | 600 | 0.37 | 0.18 | 0.48 | 0.51 |
| 11 | word2vec | 20 | 600 | 0.45 | 0.28 | 0.51 | 0.65 |
| 12 | GloVe | 20 | 600 | 0.39 | 0.19 | 0.48 | 0.49 |
| Pretrained word2vec | word2vec | - | 300 | 0.68 | 0.50 | 0.66 | 0.41 |
| Pretrained GloVe | GloVe | - | 300 | 0.59 | 0.38 | 0.56 | 0.62 |
| WordNet* | - | - | - | 0.19 | 0.39 | 0.15 | -0.04 |
| ConceptNet | - | - | - | **0.80** | **0.59** | **0.73** | 0.62 |
| Shi et al[+] | - | - | - | - | - | - | - |

\* WordNet path similarity was used in measurements.
+ The public interface created by Shi et al. (2017) does not support the retrieval of the quantitative relationships between the benchmark term pairs

## 4.3 Structural Characteristics of the TechNet

Since our focus is to provide a large-scale semantic network of terms and their semantic associations that make the best sense for engineering knowledge retrieval and inference, we choose TechNet #9 for further applications and illustrations in this paper. The relatively best TechNet (#9) consists of 4,038,924 technology-related terms in the semantic network and roughly $8.15 \times 10^{12}$ bidirectional quantified relevance values between each possible pair of terms. By contrast, WordNet contains 155,236 entities and 647,964 relations, ConceptNet contains 516,782 entities and about $1.3 \times 10^{11}$ relations, among others (Table 6). We further analysed the structural characteristics of this specific network. For simplicity, hereafter we will refer to this specific network as the "TechNet".

Table 6: Statistics of existing semantic networks

| Semantic Networks | Number of Entities | Number of Relations |
|---|---|---|
| TechNet | 4,038,924 | ~$8.5 \times 10^{12}$ |
| WordNet | 155,236 | 647,964 |
| ConceptNet | 516,782 | ~$1.3 \times 10^{11}$ |
| Pretrained word2vec | 3,000,000 | ~$4.5 \times 10^{12}$ |
| Pretrained GloVe | 417,194 | ~$1.7 \times 10^{11}$ |
| Shi et al. (2017) | 536,507 | 3,726,904 |

Note: The count statistics for the existing public semantic network datasets are from author's calculations. Shi et al. (2017) reported the number of entities and number of relations of their network.

First of all, these more than 4 million terms of the TechNet's are distributed in all technology domains as defined by the 125 3-digit patent classes in the Cooperative Patent Classification system. Across all patent classes, the distribution of terms is highly correlated (Pearson correlation coefficient = 0.976) with the distribution of patents, indicating the TechNet provides propositional and balanced coverage of the knowledge in relatively large and small domains in the total technology space. Fig. 7 reports the numbers of TechNet terms in the largest 50 technology domains by the count of patents. The distribution is skewed towards a few technology domains. The coverage of technical terms for the domains of H01-Electric Elements (890,753 terms), H04-Electric Communication Technique (960,143 terms), G06-Computing (955,223 terms), A61-Medical & Hygiene (875,240 terms) and G01-Measuring & Testing (842,668 terms) is dramatically higher than the rest. At the other extreme, 6,765 terms were found in the smallest domain G12-Instrument Details by patent count (529 patents).

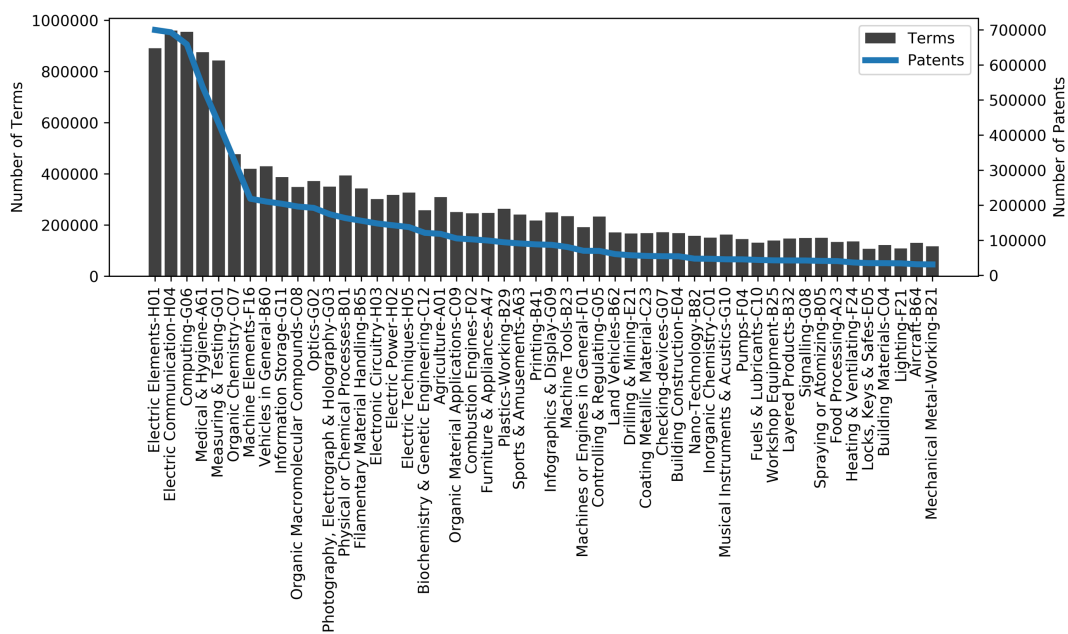

Fig. 7: Number of TechNet terms in the largest 50 technology domains by patent count

Secondly, with regard to inter-term relationships in the network, Fig. 8 shows the distribution of the relevance values of randomly picked $10^8$ pairs of terms from the TechNet, which resembles a normal distribution with a mean ($\mu$) of 0.133 and a standard deviation ($\sigma$) of 0.063. According to the distribution, more than 99.997 % of the term pairs have relevance values greater than 0. The TechNet is extremely large, dense and difficult to visualize as a network. Even if one only focuses on the strong links that has relevance values greater than three standard deviations above the mean ($\mu+3\sigma$), i.e., the top 0.15% of values in a normal distribution, the filtered network still contains about $12 \times 10^9$ links or around 6,000 strong links per term on average. The size and density of the TechNet requires efficient methods for information storage and retrieval applications.

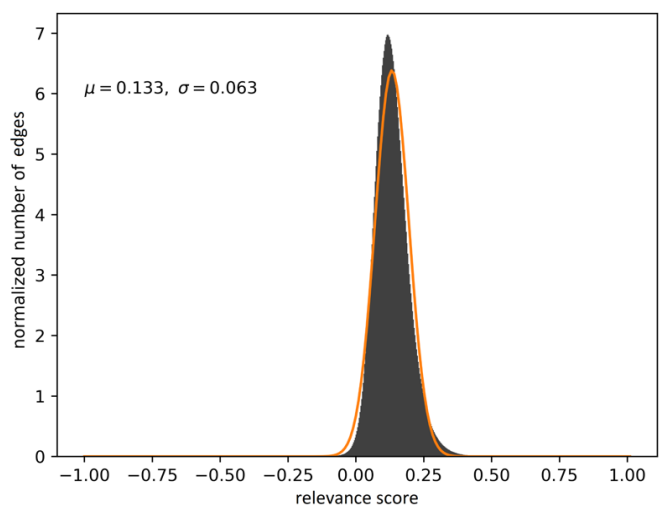

Fig. 8: The distribution of relevance scores (link weights) between randomly picked $10^8$ pairs of terms in the TechNet

## 5. Applications

The TechNet as a graph-based system of technology and engineering-related knowledge elements and their associations can serve as an infrastructure for broad uses and applications in engineering knowledge discovery and retrieval, design and innovation support and knowledge management. For example, the TechNet can be used to capture specific technology concepts from raw technical data and discover the relevant knowledge concepts around them according to semantic relations for learning and augmenting design ideation. The semantic relations also enable query prediction and expansion to make technology-related searches or knowledge discovery more intelligent. Such relational information can also aid in the search for solutions to specific engineering design problems or topics. In addition, the TechNet can be used to store, associate and organize unstructured data on technologies in image, audio or text forms for intelligent knowledge management and retrieval. Likewise, the ImageNet by Stanford University has utilized the WordNet to store, organize and retrieve image data. In this regard, the TechNet may complement the existing public semantic databases, e.g., WordNet, ConceptNet, for its strength in technology or engineering-related applications.

To enable a wide range of applications, we have developed a web interface and APIs for the public to retrieve terms and their semantic relations from the TechNet. The interface can be accessed via the URL http://www.Tech-Net.org/. The API definitions are stored in TechNet GitHub repository https://github.com/SerhadS/TechNet. The overall architecture and service framework of TechNet API is depicted in Fig. 9. We designed a Representational State Transfer (REST) web service APIs to handle basic function requests. Since it is not practical to keep the large graph database of TechNet in the server memory (over 4 million vectors of the length 600 and their pairwise relations), we designed the backend as an on-demand system where we keep the information on the storage and make use of look-up tables to call it when necessary. The main computational functions are conducted on the server side instead of the client side. As a result, the web interface is highly responsive even for mobile devices.

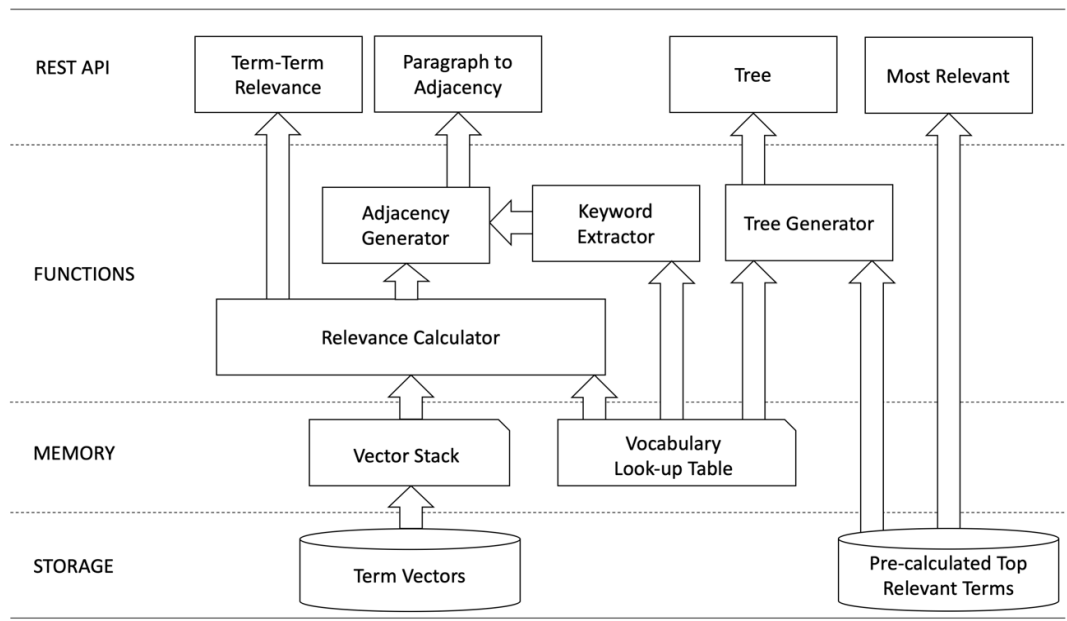

Fig. 9: Block Scheme of the TechNet API

At this moment, the interface and APIs provide four major functions. The first is to retrieve the pairwise semantic relevance between two engineering terms. For example, in Fig. 10, "autonomous vehicle" and "blind spot detecting" are related, with a semantic relevance value of 0.572. Such term-to-term relevance values can be used by researchers for their analyses.

![Term-Term Relevance: autonomous vehicle, blind spot detecting, Search!, Similarity: 0.572]

Fig. 10: Pairwise semantic relevance function in the TechNet interface

One can also use the interface or API to retrieve the most relevant terms to a term of interest. Table 7 presents the result of retrieving the 20 most relevant terms to the term "wireless charger" in the TechNet. These terms closely related to "wireless charger" represent technical concepts regarding functions, components, configurations or working mechanisms. By contrast, neither WordNet (Fellbaum, 2012; Miller et al., 1990), ConceptNet (Speer et al., 2016; Speer & Havasi, 2012; Speer & Lowry-Duda, 2017) nor the semantic network of Shi et al. (2017) contain the "wireless charger" term. In particular, we checked Google Knowledge Graph's term recommendations for "wireless charger", and the results are more related to consumer brands and products that have wireless charging capabilities (Table 7). Note that the Google Knowledge Graph is trained on Google News, Wikipedia and other layman sources of data. The TechNet appears to be more suitable for the retrieval of engineering or technical terms and their pairwise relevance. Such capability is essential for knowledge discovery in searches, recommendations, ideation, brainstorming or advisory applications.

Table 7: Top 20 most related terms to "wireless charger" in TechNet and from Google's recommendations in Google Image Search

| TechNet | | Google Image Search | |
|---|---|---|---|
| transmitter | wireless charging module | iPhone | Samsung s7 |
| transcutaneously transfer power | charging | Samsung | s6 edge |
| wireless charging | charging power | iPhone 8 | iPhone 6 |
| charger block | charging system | apple | galaxy s6 |
| wireless power | power transfer field | s7 edge | note 5 |
| maintenance charging mode | wireless power transmitter | phone | diy |
| charging power wirelessly | charging kit | car | fantasy |
| charger | battery charger | iPhone 7 | s8 plus |
| wirelessly chargeable | wireless charging field | idea | Baseus |
| full-orientation | recharge | Samsung s7 | homemade |

Our interface and API also enable retrieval of a subgraph of the TechNet that contains the technical terms from a given text, in the form of an adjacency matrix. Our web interface directly visualizes the adjacency matrix for users to easily interpret the relations among terms in the input text, and also provides the matrix data via a CSV file download for one to conduct their own analysis. The example in Fig. 11 presents the term adjacency matrix based on the short text on "radio technology" from Wikipedia: "Radio is the technology of using radio waves to carry information, such as sound and images, by systematically modulating properties of electromagnetic waves." Such retrieval includes not only the technical concepts but also their pairwise relevance together as a subgraph in the total TechNet and can be useful for a wide range of text analyses.

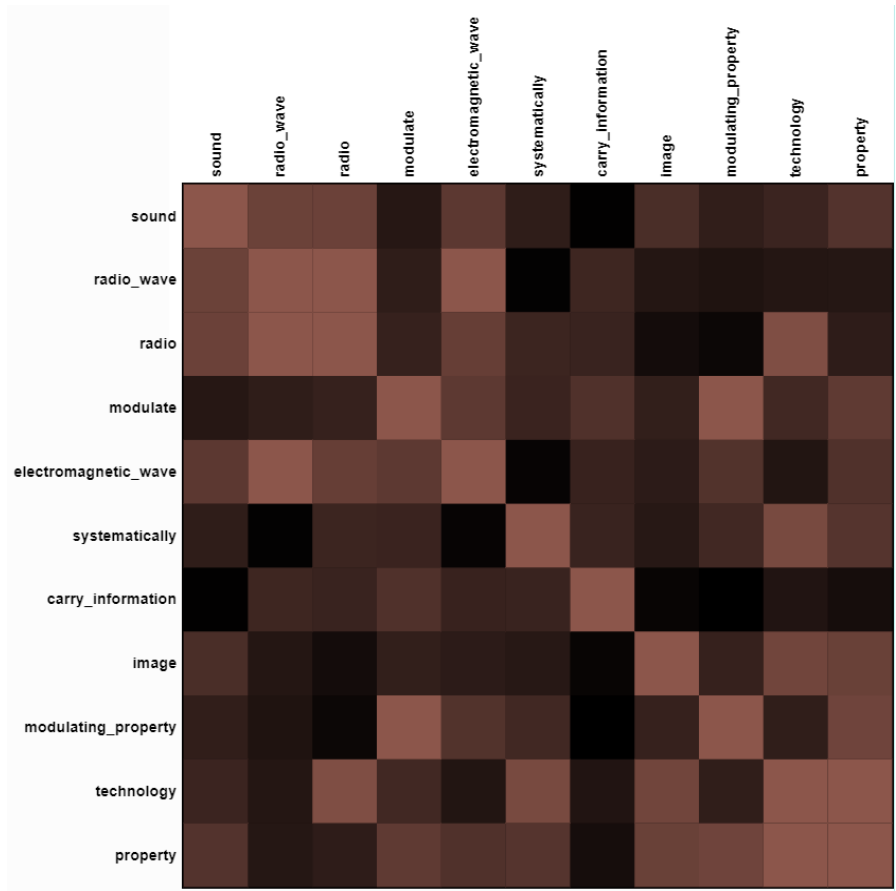

Fig. 11: Color coded visualization of adjacency matrix of the key terms contained in the Wikipedia entry for "radio technology". Lighter colors stand for higher relevancy.

In addition, the interface also allows users to manually discover the most relevant terms from a user-defined root term through a tree-expansion graph search from the root term. Fig. 12 displays the term tree expanding from the root "flying car" concept with a depth of 3 layers and a breadth of 3 branches in each layer. Alternatively, in the TechNet interface, one can manually and heuristically decide the expansion branches from each term and the layers for expansion. These surrounding concepts in the tree may provide a medium to quickly explore not the closest, but still relevant concepts for the focal concept. Such a function might facilitate related divergent thinking in engineering design ideation and brainstorming.

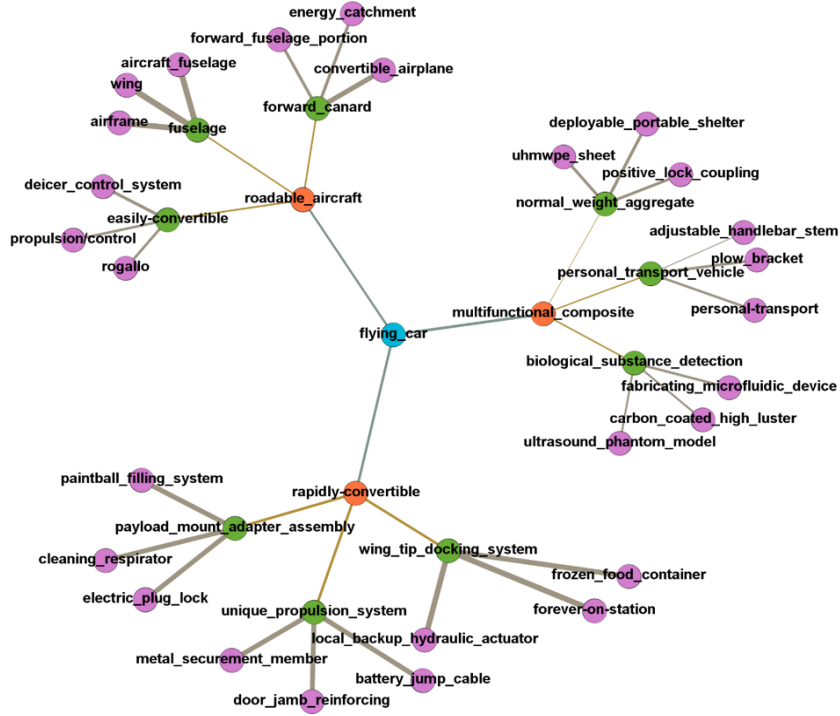

Fig. 12: A tree of concepts around "flying car" with breadth and depth of 3

Interested readers may test and explore the forgoing TechNet-enabled analytics at http://www.Tech-Net.org for their specific interests. In the meantime, the applications of the TechNet are not limited in these ones presented in this paper. We will add new functions and also invite researchers from different fields to develop broad applications of the TechNet.

## 6. Discussion and Concluding Remarks

Our approach to construct the TechNet, to the best of our knowledge, is the first to train the recently-emerged word embedding models on the complete USPTO patent text database to construct a comprehensive semantic network of engineering concepts with technically-meaningful semantic associations. By contrast, in the prior literature, word embedding models have been only used in non-engineering text analytics; and, patent texts have only been analyzed at a small sample scale (rather than the total patent database). Our work also necessarily included the evaluation and selection of the trained networks (arising from different hyperparameters in the training process), and for this purpose, we curated for the first time a Technical Term Relevance (TTR) benchmark dataset based on the evaluation of experienced engineers in the context of engineering. Despite the existence of benchmark tasks and datasets for NLP in non-engineering contexts, they were not created and thus unsuitable for evaluating semantic networks for engineering knowledge retrieval and inference.

In turn, the novel combination of the total patent database (as the data source), word-embedding models (as the method), and the new evaluation benchmark in the context of engineering (as the application context) has led to the first-of-its-kind TechNet. We have been able to identify the TechNets (especially the ones based on word2vec) that outperform existing public semantic networks (e.g., WordNet, ConceptNet, Google Knowledge Graph) for knowledge retrieval and inference tasks relevant to engineering and technology, even though the present study only mined patent titles and abstracts, employed two word embedding algorithms and a small set of hyperparameter values, and curated a small TTR dataset. Therefore, there are great potentials to derive even better-performing TechNets in future research via expanding the training database, fine tuning the construction procedures, training settings and the benchmark datasets, as well as exploring alternative techniques.

First of all, mining technical description texts of patent documents (and also other patent databases than the USPTO database) might further increase the coverage of the engineering lexicons and enrich the word embeddings training. Secondly, a wider set of alternative term extraction techniques can be explored, tested and compared to improve the corpus. For example, our current term extraction procedure involves a considerable manual effort for detecting noisy terms. Automation of this step would allow a wider exploration of TechNet constructions. The third is to explore more advanced word embedding algorithms beyond word2vec and GloVe and a wider range of hyperparameter values for training. Given the trained term vectors, metrics other than cosine similarity to associate them should be explored and tested. Furthermore, the TTR benchmark dataset can be further improved by including more diverse term pairs and engaging more human expert evaluators. On top of these, we will continually improve the functions and features of the TechNet web portal (www.tech-net.org) and APIs (https://github.com/SerhadS/TechNet) for public users to explore applications of TechNet.

The TechNet fills the gap of a large-scale technology semantic network to augment knowledge-based intelligence for engineering and technology-related applications. Moving forward, instead of standing-alone applications, TechNet could also be integrated with the existing general-knowledge semantic networks to empower them for technology-related text analysis and artificial intelligence applications. The vectorized structure of the TechNet would make the integration easy. Particularly, in our evaluation against the TTR benchmark (Table 5, right most column), the ConceptNet corresponded the best among all general-knowledge semantic networks and even better than the GloVe-trained candidate TechNets, whereas it presents significantly lower coverage of technical terms than the TechNets. These findings suggest the prospects to integrate the TechNet (with superior performance in the specific technical context) with ConceptNet.

In sum, this research is only the first step to build the technology semantic network. As new technologies continue to emerge and the patent database continues to grow, the TechNet will need to be regularly updated and scaled up by further training with new patent data. Additionally, the advancement in data science and, particularly, NLP techniques offers new and better means to construct the corpus and word embeddings and fine-tune the TechNet. In turn, the TechNet will serve as an infrastructure to enable the development of many new applications of artificial intelligence for engineering design, knowledge management, and technology innovation.